\documentclass[twocolumn,twoside,slac_two]{revtex4}
\usepackage{graphicx}
\usepackage{fancyhdr}
\begin{document}

\title{Lepton Flavor Violating Decays -- Review \& Outlook}


\author{Toshinori Mori}
\affiliation{International Center for Elementary Particle Physics, The University of Tokyo \\ 
7-3-1 Hongo, Bunkyo-ku, Tokyo 113-0033, Japan}

\begin{abstract}
Here I review the status and prospects of experimental investigations 
into lepton flavor violation (LFV) in charged leptons. 
Rare LFV processes are naturally expected to occur through loops of 
TeV scale particles predicted by supersymmetric theories or 
other models beyond the Standard Model.  
In contrast to physics of quark flavors that is 
dominated by the Cabibbo-Kobayashi-Maskawa matrix, 
LFV in charged leptons is a definitive signal of new physics.
Currently active researches are rare tau decay searches 
at the B factories.  
The MEG experiment will soon start a sensitive search for 
the LFV muon decay, $\mu\rightarrow$e$\gamma$.  
Prospects for searches at the LHC, 
a possibility of a fixed target LFV experiment with high energy muons, 
and a sensitivity of leptonic kaon decays to LFV 
are also briefly discussed. 
\end{abstract}

\maketitle

\thispagestyle{fancy}


\section{Why Lepton Flavor Violation?}

Flavor violation or mixing among quarks has been known for many years and 
is beautifully described by the Cabibbo-Kobayashi-Maskawa matrix 
in the Standard Model.  
On the other hand, the discovery of flavor violation or oscillation among 
neutrinos came as a big surprise and provides a possible hint of 
new physics beyond the Standard Model.  
Now lepton flavor violation (LFV) among charged leptons, 
which has never been observed, 
is attracting a great deal of attention, 
because its observation is highly expected by many of the well motivated theories and 
would undisputedly establish a breakdown of the Standard Model.  

\begin{figure}[h]
\centering
\includegraphics[width=75mm]{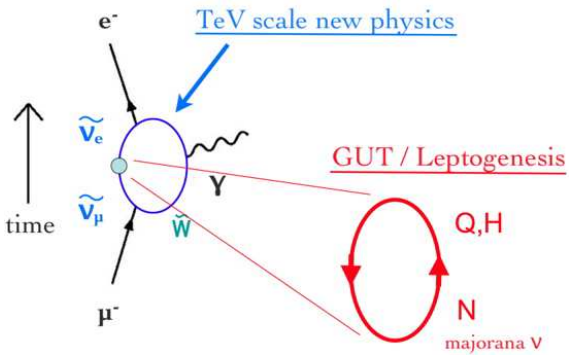}
\caption{A possible origin of LFV processes 
($\mu\rightarrow\mathrm{e}\gamma$ in this example).} \label{fig:megloops}
\end{figure}

An example of such LFV processes is schematically indicated in 
Figure~\ref{fig:megloops}.  
LFV is expected to occur in the loops of new physics processes at TeV scale 
such as supersymmetry or extra dimensions. 
Therefore the discovery of such a LFV process is of similar significance 
to that of the LHC.  

On the other hand, the source of LFV originates from much higher energy scale 
governed by grand unification theories (GUT)~\cite{ref:barbieri} 
or seesaw models that predict heavy majorana neutrinos 
to derive tiny neutrino masses~\cite{ref:hisano}, 
as indicated by the red loop in the Figure.  
Therefore the discovery and measurement of LFV processes 
could also provide hints of physics at extremely high energy scale, 
which would not be accessible even at the LHC.  

In this article the present and future {\it experimental} researches 
on LFV in charged leptons are reviewed.

\section{LFV Tau Decays}

\begin{figure}[h]
\centering
\includegraphics[width=60mm]{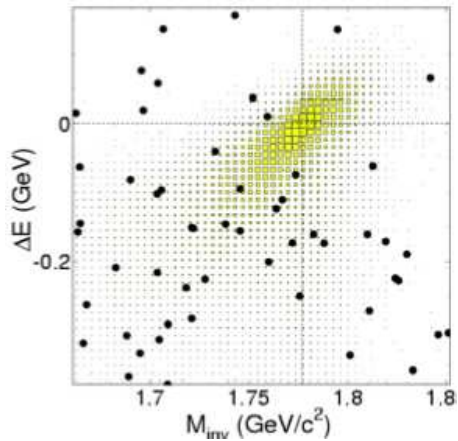}
\caption{A result of $\tau\rightarrow\mu\gamma$ search by Belle 
for 86.7~fb$^{-1}$ data, giving a 90\% C.L. limit of $3.9\times 10^{-7}$.  
Shown in yellow is the expected distribution for the signal 
while the points are the data.  The background events are mostly $\tau$-pair events 
with large initial state radiation.} \label{fig:belle}
\end{figure}

Currently most actively studied LFV processes are the rare $\tau$ decays.  
$\tau$-pairs are abundantly produced at the B factories where the $\tau$-pair production 
cross section is as large as that of $B\bar{B}$.  
The two B factory experiments, Belle and BaBar, have accumulated 
more than $7.5\times 10^8$ $\tau$-pairs altogether.  
$\tau$-pair events are selected and tagged by one of the $\tau$s that decayed in the 
normal way and the $\tau$s on the other side are searched for LFV decays.  
A result of such analyses is shown in Figure~\ref{fig:belle}. 
As can be seen from this Figure, many of the searches are already beginning 
to be limited by background events.  
A more detailed description of various searches for LFV $\tau$ decays is 
given by Dr. H.~Kakuno in this conference~\cite{ref:kakuno}.  
They have made an impressive improvement on most of the LFV modes 
of the $\tau$ decays, 
though any of them has not been discovered yet.  
The 90\% C.L. upper limits on their branching ratios are now in the order of $10^{-7}$, 
an order of magnitude improvement over the previous experiments 
(mostly by CLEO)~\cite{ref:inami}.  

These limits strongly constrain new physics models such as supersymmetry, 
especially for a large $\tan\beta$ region and also for Higgs-mediated LFV vertices.  

For the future these limits should improve as the B factories continue to accumulate more data, 
but the improvements would be slow due to the background especially for some modes 
such as $\tau\rightarrow\mu\gamma$/e$\gamma$, $\tau\rightarrow\mu\eta$, etc.  
A Super B Factory with 5--10~ab$^{-1}$ would bring them into the $10^{-8}$ region, 
possibly even to $10^{-9}$ for some modes, but claiming a discovery would be much 
harder with the existing background events.  

Preparatory studies on LFV $\tau$ decays are being conducted 
by the LHC experiments~\cite{ref:giffels}.  
During their initial low luminosity runs (10--30~fb$^{-1}$/year) for 2008--2009, 
a search for $\tau\rightarrow 3\mu$ may be possible.  
$\tau$s are produced by the decays of W and Z (an order of $10^8$ $\tau$s per year) 
and B ($10^{11}$ $\tau$s).  In order to be competitive, 
more abundant $\tau$s coming from D$_s$ decays ($10^{12}$ $\tau$s/year) 
are being investigated.   A preliminary study indicates that 
efficient triggering of these $\tau$s would be very difficult
because of the low $p_t$ (less than a few GeV) of the decay muons.

\section{Muon to Tau Conversion} 

In a model independent approach, 
although many of the effective LFV four-fermion couplings involving $\tau$s 
are already strongly constrained by the LFV $\tau$ decay searches, 
there are some couplings that are still only loosely constrained~\cite{ref:sher}.  
For example, couplings involving heavy quarks cannot be studied 
by the $\tau$ decays, 
because $\tau$s cannot decay into heavy flavor hadrons.  
Such couplings may be studied by using high intensity muon beam 
on a fixed target and looking for 
a conversion from muons to $\tau$s~\cite{ref:mutau_conv}.  

In SUSY models such a muon to $\tau$ conversion could be enhanced 
by Higgs mediation~\cite{ref:kanemura}.  
With the present limits on $\tau\rightarrow\mu\eta$, $3\mu$, 
a maximum of $\approx 100\rho$ conversion events are possible 
for $10^{20}$ incident 50~GeV muons with a target thickness of $\rho$ g/cm$^3$, 
and more events for higher energy muons.  In fact, above 60~GeV, subprocesses 
involving b-quark component inside the target dominate and significantly 
increase the cross section.  
Such a high intensity, high energy beam may be available in the future 
at a neutrino factory or a muon collider, or, for an electron to $\tau$ conversion, 
the International Linear Collider (ILC).  

A preliminary study with a conceptual experimental design 
indicates that designing an experiment for a muon rate of $3\times 10^{11}$/sec/m$^2$ 
would not be totally unfeasible but challenging and expensive~\cite{ref:mutau_conv}.

\section{Leptonic Kaon Decays}

As Dr.A.~Ceccucci also pointed out in this conference~\cite{ref:ceccucci}, 
the ratio of the leptonic decays of kaons, 
$R_K = (K\rightarrow \mathrm{e}\nu) / (K\rightarrow \mu\nu)$, 
with theoretical uncertainties canceling in the ratio, 
was recently shown to be very sensitive to LFV couplings~\cite{ref:masiero}.   
In SUSY models the LFV decay of kaon into an electron or a muon with 
a $\tau$ neutrino can be strongly enhanced by charged Higgs mediation, 
while LF conserving SUSY effects do not contribute much.  
In fact a measurement with a 1\% precision will lead to strong constraints 
on LFV vertices corresponding to upper limits on the LFV $\tau$ decays, 
$Br(\tau\rightarrow \mathrm{e}\eta) < 10^{-10}$ or 
$Br(\tau\rightarrow \mathrm{e}\gamma) < 10^{-11}$,  
i.e. exceeding even the Super B Factory sensitivity. 

The NA48/2 experiment at CERN recently presented their preliminary result 
of $R_K = 2.416\pm 0.043\pm 0.024$ using their data taken in 2003~\cite{ref:NA48} 
for the Standard Model prediction of $R_K = 2.472\pm 0.001$, 
indicating a possible one-sigma deviation.  
The data statistics will be at least doubled with their 2004 data.  
The NA48/2 collaboration now plans to take more data in 2007 
to improve their measurements.  

The leptonic $\pi$ decay ratio, 
$R_{\pi} = (\pi\rightarrow \mathrm{e}\nu) / (\pi\rightarrow \mu\nu)$, 
also has a LFV sensitivity which is much smaller than $R_K$ 
by $(m_{\pi}/m_K)^4$.  
Two experimental programs, one at TRIUMF and the other at PSI, 
have been approved to investigate possibilities to improve the present precision of 
0.3\%.

\section{LFV Muon Decays} 

Well motivated theories of supersymmetric grand unification (SUSY GUT) 
are very strongly constrained by the upper limits of the branching ratios of 
the LFV muon decays, 
$\mu\rightarrow \mathrm{e}\gamma$, and 
$\mu\rightarrow \mathrm{e}$ conversion on a nucleus.   
Therefore any improvement in experimental sensitivities to these decays 
could possibly lead to a discovery of SUSY GUT signals.  

Assuming the standard Gauge (photon) mediated couplings, 
physics sensitivity of $\mu\rightarrow \mathrm{e}\gamma$ is 
300--400 times higher than that of  $\mu\rightarrow \mathrm{e}$ conversion, 
depending on the conversion target nuclei.  
Thus a branching ratio sensitivity of $10^{-13}$ for $\mu\rightarrow \mathrm{e}\gamma$ 
corresponds to a $\approx 3\times 10^{-16}$  
$\mu\rightarrow \mathrm{e}$ conversion sensitivity.  

While an experimental search for $\mu\rightarrow \mathrm{e}\gamma$ is 
mainly limited by accidental coincidence of background events, 
prompt backgrounds from muon beams are major obstacles 
in a $\mu\rightarrow \mathrm{e}$ conversion experiment.  
Consequently a DC muon beam is best suited to 
a $\mu\rightarrow \mathrm{e}\gamma$ search 
to minimize accidental overlap.  On the other hand, a pulsed beam is utilized 
to reduce prompt background in a $\mu\rightarrow \mathrm{e}$ conversion search.  

There is another interesting decay mode, $\mu\rightarrow$3e, 
which, assuming the Gauge-mediated couplings, has a roughly 100 times 
smaller branching ratio than $\mu\rightarrow \mathrm{e}\gamma$.  
Its major experimental limits come from accidental overlaps of 
background events.  
At the moment there is no experimental program, present or planned, 
for this decay mode.  

In the following the status and prospects of the experimental programs to 
search for $\mu\rightarrow \mathrm{e}\gamma$ and 
$\mu\rightarrow \mathrm{e}$ conversion 
are briefly summarized.

\subsection{Conversion to Electron on Nucleus}

\begin{figure}[h]
\centering
\includegraphics[width=80mm]{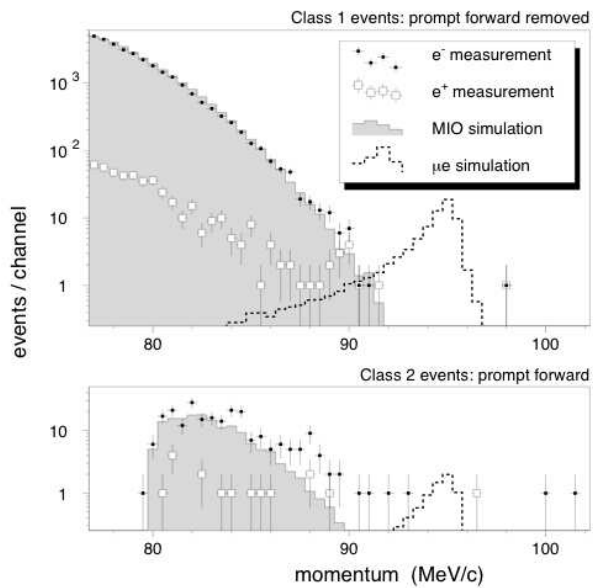}
\caption{The final result of the SINDRUN-II experiment for the 
gold target.} \label{fig:sindrum2}
\end{figure}

The present best limit on the $\mu\rightarrow \mathrm{e}$ conversion 
is given by the SINDRUM-II experiment at PSI.  
Figure~\ref{fig:sindrum2} shows their final result on gold target, 
giving a 90\% C.L. limit of $7\times 10^{-13}$~\cite{ref:sindrum2}.   
As seen in the bottom figure, a class of events in coincidence with 
the 20~nsec beam pulse contain a lot more background.  

The MECO experiment, proposed at BNL to achieve a $10^{-16}$ 
sensitivity by using a graded field solenoid to collect as much 
pions as possible at the production target, 
was unfortunately cancelled by the U.S. funding agency in 2005.  

\begin{figure}[h]
\centering
\includegraphics[width=60mm]{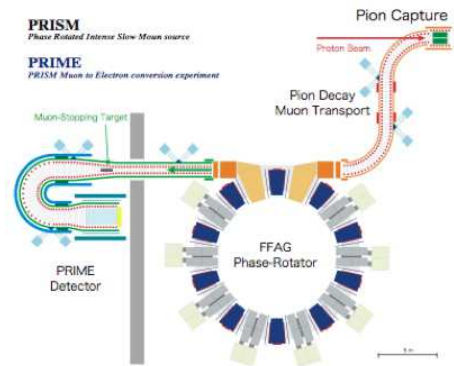}
\caption{A possible layout of the PRISM/PRIME facility proposed 
at the J-PARC 50~GeV proton ring.} \label{fig:prism}
\end{figure}

PRISM (Phase-Rotated Intense Slow Muons) is an ambitious project 
to produce high intensity negative muon beam with narrow energy spread 
and much less $\pi$ contamination, proposed to be built at the J-PARC 
50~GeV proton ring that is currently under construction 
at Tokai, Japan~\cite{ref:prism}.  
Its schematic layout is shown, together with the proposed 
$\mu\rightarrow \mathrm{e}$ conversion experiment PRIME, 
in Figure~\ref{fig:prism}.  
A Fixed-Field Alternating Gradient synchrotron (FFAG) is used to 
carry out ``phase rotation," i.e. a conversion of 
an original short pulse beam with wide momentum spread ($\pm 30$~\%) 
into a long pulse beam with narrow momentum spread ($\pm 3$~\%)
by strong RF field.  
After 5 turns in the FFAG ring for the phase rotation, 
pions in the beam all decay out. 
Given $10^{14}$ protons/sec from the J-PARC ring, the PRISM 
facility should be able to provide $10^{11}$ - $10^{12}$ muons/sec, 
which might enable a $\mu\rightarrow \mathrm{e}$ conversion sensitivity  
down to $10^{-18}$.  
There are still several R\&D items to study: for example, 
low energy pion production and capture system, and 
injection/extraction of muons into/from the FFAG ring. 
A real-size FFAG ring is being constructed at Osaka University for R\&D studies. 
The schedule of the project is unknown as it is not funded yet.  

In conclusion there is no active or approved experiment for 
a $\mu\rightarrow \mathrm{e}$ conversion search at the moment.  
%It is therefore hoped that a modest size experiment to reach $10^{14}$--$10^{-16}$ 
%might be designed and quickly carried out at an existing facility 
%by international efforts, 
%before the ambitious PRISM project could be eventually realized.  

\begin{figure*}[t]
\centering
\includegraphics[width=135mm]{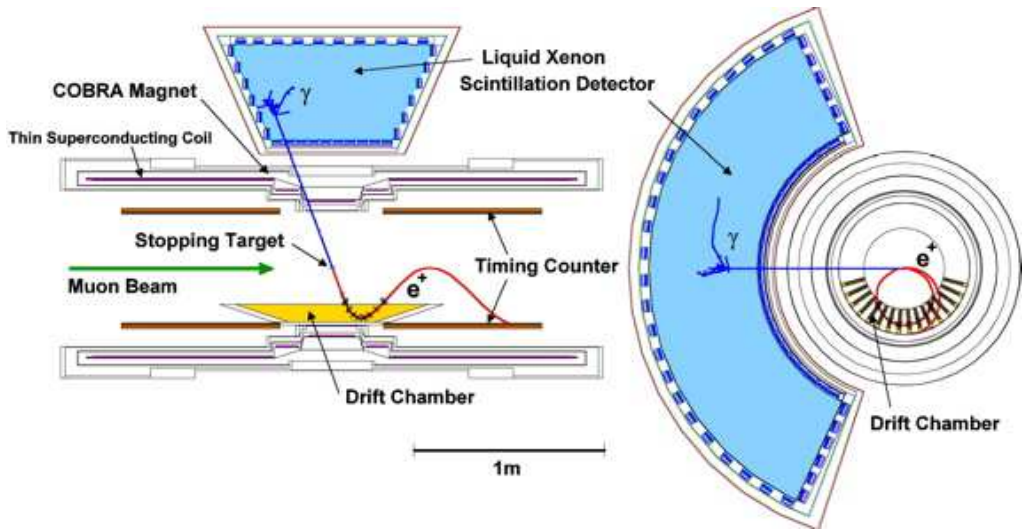}
\caption{A schematic view of the MEG experiment.} \label{fig:meg}
\end{figure*}

\subsection{MEG Experiment}

The MEG experiment~\cite{ref:MEG}, 
a $\mu\rightarrow \mathrm{e}\gamma$ search experiment, 
currently being prepared at the Paul Scherrer Institute (PSI) in Switzerland, 
was proposed by the Japanese physicists and has since evolved to 
an international collaboration among Japan, Italy, Switzerland, Russia, 
and the U.S.A.  
The experiment is scheduled to be ready for physics run towards the end of 2006 
and aims at a sensitivity of $10^{-13}$, two orders of magnitude below 
the present limit of $1.2\times 10^{-11}$~\cite{ref:MEGA}.  

\begin{figure}[hb]
\centering
\includegraphics[width=75mm]{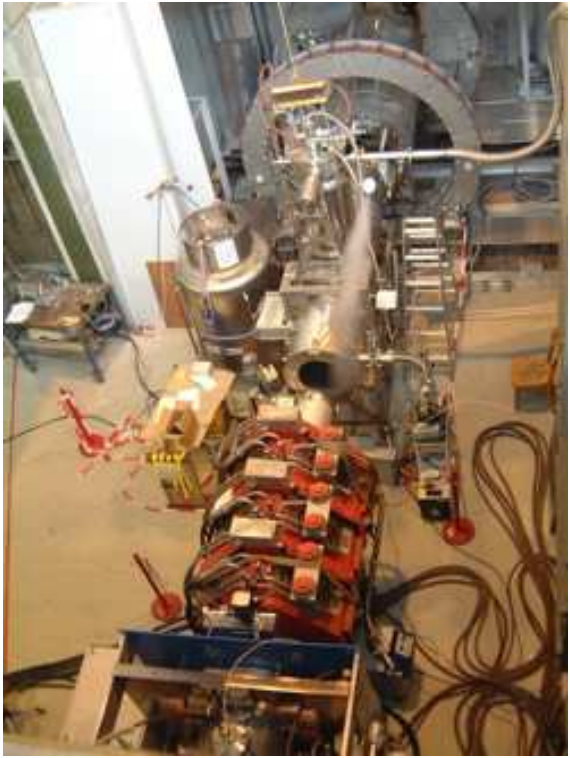} 
\caption{The $\pi$E5 beam line at PSI.  From upstream (bottom) to 
downstream (top) are 
the DC separator, the quadrupole magnets (red), 
the superconducting muon transport magnet, and 
the COBRA magnet with ring-shaped compensation coils.} \label{fig:meg_beamline}
\end{figure}

The experimental set-up is schematically shown 
in Fig.~\ref{fig:meg}.  
A DC surface muon beam of a few times $10^{7}$/sec is focused and stopped 
in a thin plastic target.  
Gamma rays from $\mu\rightarrow \mathrm{e}\gamma$ decays are 
measured by liquid xenon scintillation detector 
located just outside a very thin solenoidal magnet called COBRA.  
Positrons are tracked by low material drift chambers 
inside COBRA which provides specially graded 
magnetic field.  Their timings are measured by plastic scintillation 
counters in the second turn of their trajectories. 

The unprecedented sensitivity of the MEG experiment 
has been made possible by the three key components: 
(1) the highest intensity DC surface muon beam available at PSI; 
(2) a specially designed COBRA positron spectrometer with graded magnetic field; and 
(3) an innovative liquid xenon scintillation gamma ray detector. 
These key components are described in some detail 
in the following.  

The 590MeV proton cyclotron at PSI, 
constantly operating with a beam current exceeding 1.8mA 
and a total beam power of more than 1~MW, 
is able to produce the highest intensity DC muon beam 
in the world.  This is the best and only place suitable 
for a $\mu\rightarrow \mathrm{e}\gamma$ experiment.  

The $\pi$E5 beam line for the MEG experiment 
is shown in Figure~\ref{fig:meg_beamline}.  
The superconducting solenoidal magnet is used to transport 
and focus the beam onto a small target (a few cm$\phi$). 
A DC separator is placed to reject unwanted positrons in the beam. 
A muon rate of $10^8$/sec has been already demonstrated so that 
more than $10^{15}$ stopped muons per year is reasonably expected.  

The COBRA (COnstant Bending RAdius) positron spectrometer~\cite{ref:cobra} 
consists of a superconducting solenoidal magnet 
designed to form a special graded magnetic field 
(1.27~T at the center and 0.49~T at the both ends), 
in which positrons with the same absolute momenta follow 
trajectories with a constant projected bending radius, 
independent of the emission angles over a wide angular range.  
This allows to sharply discriminate high momentum signal 
positrons ($p = m_{\mu}/2 = 52.8$MeV/$c$) out of $10^7$--$10^8$ Michel 
positrons emitted every second from the target.  
Only high momentum positrons enter the drift chamber volumes. 
The graded field also helps to sweep away curling tracks 
quickly out of the tracking volume, 
thereby reducing accidental pile-up of the Michel positrons. 

High strength Al stabilized conductor is used to make the magnet 
as thin as 0.197$X_0$, so that 85\% of 52.8MeV/$c$ gamma rays 
traverse the magnet without interaction before entering 
the $\gamma$ ray detector placed outside the magnet.  
A He-free, simple and easy operation of the magnet is 
realized with a GM refrigerator.  
As the COBRA magnet does not have a return yoke, 
a pair of compensation coils (seen as large rings in 
Figure~\ref{fig:meg_beamline}) suppress the stray magnetic field 
below 50~Gauss in the vicinity of the gamma ray detector, 
so that the photomultiplier tubes can operate. 

\begin{figure}[h]
\centering
\includegraphics[width=80mm]{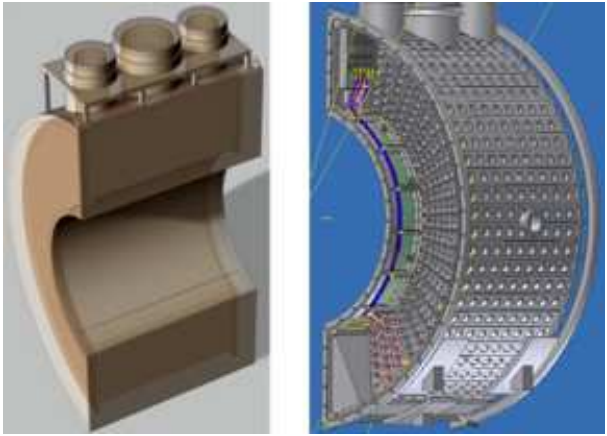}
\caption{The liquid xenon $\gamma$ ray detector.  The cryostat (left) and the 
support structure for the photomultiplier tubes inside the cryostat (right).} \label{fig:lxe}
\end{figure}

An innovative liquid xenon (LXe) scintillation detector was specially 
devised for this experiment to 
make very precise measurements of energy, position 
and timing of gamma rays.  
The detector holds an active LXe volume of 800~$\ell$.  
Scintillation light emitted inside LXe are viewed from all sides 
by approximately 850 photomultiplier tubes (PMTs) that are immersed 
in LXe in order to maximize direct light collection.  

High light yield of LXe (roughly 75\% of NaI) 
and its uniformity are necessary ingredients for good energy resolution.  
A scintillation pulse from xenon is very fast and has a short tail, 
thereby minimizing the pile-up problem.  Distributions of the PMT 
outputs enable a measurement of the incident position of the gamma ray 
with a few mm accuracy.  The position of the conversion point is 
also estimated with an accuracy that corresponds to a timing 
resolution of about 50~psec.  

Various studies were carried out using a 100~$\ell$ prototype detector 
in order to gain practical experiences 
in operating such a new device and to prove its excellent performance. 
PMTs that work at the LXe temperature (-110$^\circ$C), 
high power pulse tube refrigerators for LXe, a liquid phase purification 
method to remove possible impurities that absorb scintillation light, 
a calibration method using wire $\alpha$ sources, etc. 
have been developed~\cite{ref:LXe}. 
Several gamma ray beam tests demonstrated its excellent performance 
which is necessary to achieve the sensitivity goal of the experiment.  

The MEG detectors are currently being constructed and are scheduled to be ready 
later in the year 2006.  It is expected to take $\approx 2$ years with 
a muon beam of a few$\times 10^7$/sec to reach a 90\% C.L. sensitivity of 
$1\times 10^{-13}$ with an expected background of 0.5 events.

\section{Conclusion}

LFV in charged leptons is a clean and clear signal of new physics 
beyond the Standard Model.  
It not only evidences new physics at TeV scale, 
such as supersymmetry or extra dimensions, 
but also provides hints of physics at extremely high energies, 
such as grand unification of forces that might have triggered 
inflation of the universe, 
%by causing vacuum phase transition, 
or heavy majorana neutrinos that might be the origin of matter. 
%via leptogenesis.  

The B factory experiments have greatly improved and will further improve 
the limits on the LFV $\tau$ decays 
but are now starting to suffer background events.  
In parallel interesting limits may be provided by the measurement of 
the leptonic kaon decays and the LHC experiments could also contribute 
in their initial low luminosity phase.  

In the absence of funded projects for $\mu\rightarrow \mathrm{e}$ conversion searches, 
it is hoped that a sensible experiment be designed and started by international 
collaboration, before the ambitious PRISM facility may be eventually realized.  

\begin{figure}[hb]
\centering
\includegraphics[width=75mm]{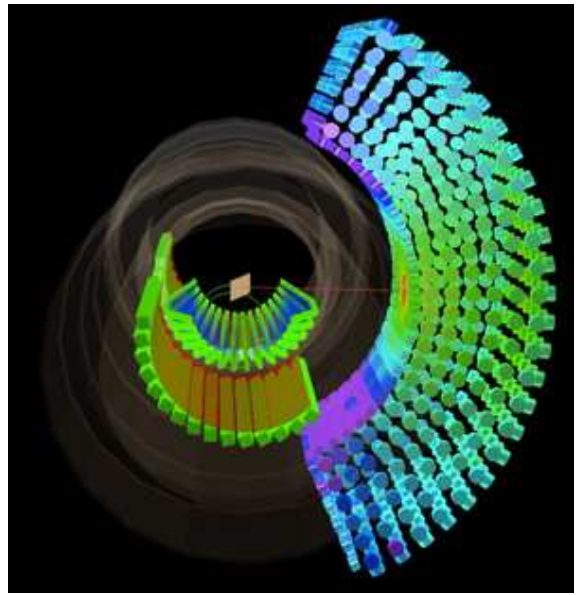}
\caption{A simulated $\mu\rightarrow\mathrm{e}\gamma$ event 
detected by the MEG detector.} \label{fig:meg_event}
\end{figure}

Finally, 
the MEG experiment is expected to be ready toward the end of 2006 
and could come across such a charming event 
as shown in Figure~\ref{fig:meg_event} at any time during the next few years.  
So stay tuned.

\bigskip % extra skip inserted
\begin{acknowledgments} 
The author thanks K.~Inami of the Belle Collaboration for useful 
discussion on the rare tau decay searches at the B factories.  
This work was supported by MEXT Grant-in-Aid for Scientific Research on 
Priority Areas.  
\end{acknowledgments}

\bigskip % extra skip inserted

\end{document}